\documentstyle[preprint,aps,epsfig,tighten]{revtex}

\begin{document}
\title{On the structure of the scalar and $D$ mesons}
\author{J. Vijande, F. Fern\'{a}ndez, A. Valcarce}
\address{Grupo de F\' \i sica Nuclear,\\
Universidad de Salamanca, E-37008 Salamanca, Spain}
\maketitle

\begin{abstract}
The ${q}\overline{q}$ spectrum is studied within a
chiral constituent quark model. It provides with
a good fit of the available experimental data from 
light (vector and pseudoscalar) to heavy mesons. 
The new $D$ states measured at different factories are studied.
The $0^{++}$ light mesons are analyzed as $q\bar q$ pairs or
tetraquark structures.
\end{abstract}

\section{Introduction}

The continuously increasing huge amount of data and its simplicity
have converted meson spectroscopy in an ideal system to learn about
the properties of QCD. Since Gell-Man conjecture, most of the 
meson experimental data were classified as $q\overline{q}$ 
states according to $SU(N)$ irreducible representations. 
Nevertheless a number of interesting issues remains still open
as for example the understanding of some data recently obtained on the 
B factories or the structure of the scalar mesons.
The recently measured $D_s$ states are hardly accommodated in a
pure $q\bar q$ description. They present a mass much lower than the
naive prediction of constituent quark models. Besides,
the underlying structure of the scalar mesons 
is still not well established theoretically. 
In a pure $q\overline{q}$ scheme the light scalars could be identified
with the isoscalars $f_0(600)$ and $f_0(980)$, 
the isodoublet $\kappa(900)$, and the isovector $a_0(980)$,
constituting a $SU(3)$ flavor nonet.
However such identification immediately faces
difficulties to explain for example (i) the $f_0(980)$
and $a_0(980)$ mass degeneracy, (ii) why $a_{0}(980)$ and $f_{0}(600)$ 
have so a different mass?, (iii) the similar branching ratios 
of the $J/\psi \rightarrow
f_0(980)\phi $ and $J/\psi \rightarrow f_0(980)\omega$ decays which
clearly indicates the existence of strange and nonstrange content in 
the $f_0(980)$.

The theoretical tools to determine the properties of mesons are based to a 
large extent on phenomenological models.
The study of charmonium and bottomonium made clear that heavy-quark systems 
are properly described by nonrelativistic potential models reflecting 
the dynamics expected from QCD \cite{eich}. 
For heavy quarks to leading order in $v^2$ it has been demonstrated
that the interaction can be derived from the theory \cite{yndy}. 
The light meson sector has been studied by means of
constituent quark models, where quarks are dressed with a
phenomenological mass and bound in a nonrelativistic potential, usually
a harmonic oscillator \cite{isgu}.
Quite surprisingly a large number of properties of hadrons could be
reproduced in this way \cite{godf}. In this talk we present the 
meson spectra obtained by means of a chiral constituent quark model
in a trial to interpret some of the still unclear experimental data
in the light scalar and $D$ meson sectors.

\section{SU(3) chiral constituent quark model}

Since the origin of the quark model hadrons have been considered to be
built by constituent (massive) quarks. Nowadays it is widely
recognized that the constituent quark mass
appears because of the spontaneous
breaking of the original $SU(3)_{L}\otimes SU(3)_{R}$ chiral symmetry at
some momentum scale. In this domain a simple 
Lagrangian invariant under the chiral 
transformation can be derived as \cite{diak}
\begin{equation}
L=\overline{\psi}(i \gamma^\mu \partial_\mu -MU^{\gamma _{5}})\psi
\label{eq1}
\end{equation}
where $U^{\gamma _{5}}=\exp (i\pi ^{a}\lambda ^{a}\gamma
_{5}/f_{\pi })$. $\pi ^{a}$ denotes the pseudoscalar fields $(\vec{\pi }
,K_{i},\eta_8)$ with i=1,...,4, and $M$ is the constituent quark mass. An
expression of the constituent quark mass can be obtained from the theory,
but it also can be parametrized as $M(q^{2})=m_{q}F(q^{2})$ with
\begin{equation}
F(q^{2})=\left[ \frac{\Lambda^{2}}{\Lambda^{2}+q^{2}} \right] ^{\frac{1}{2}}
\end{equation}
where $\Lambda$ determines the scale at which chiral symmetry is
broken. Once a constituent quark mass is generated such particles have 
to interact through Goldstone modes. Whereas the Lagrangian 
$\overline{\psi} (i\gamma^\mu \partial_\mu -M)\psi$ is not invariant under
chiral rotations, the one of Eq. (\ref{eq1})
is invariant since the rotation
of the quark fields can be compensated renaming the bosons fields. 
$U^{\gamma _{5}}$ can be expanded in terms of boson fields as,
\begin{equation}
U^{\gamma _{5}}=1+\frac{i}{f_{\pi }}\gamma ^{5}\lambda ^{a}\pi ^{a}-\frac{1}{%
2f_{\pi }^{2}}\pi ^{a}\pi ^{a}+...
\end{equation}
The first term generates the constituent quark mass and the second one gives
rise to a one-boson exchange interaction between quarks. The main
contribution of the third term comes from the two-pion exchange which will
be simulated by means of the one-sigma exchange potential. Based on the
nonrelativistic reduction of the former Lagrangian, one obtains an
interaction between quarks mediated by the exchange 
of different Goldstone bosons. Among them, the one-pion, one-kaon 
and one-eta exchanges contribute with a central and a tensor 
interaction, while the one-sigma exchange gives a central and 
a spin-orbit term. Explicit expressions of these potentials can 
be found elsewhere \cite{vij1}. The parameters of the Goldstone 
boson exchange interaction are fixed 
assuming that $SU(3)$ flavor-symmetry is exact,
only broken by the different mass of the strange quark. 
In the heavy quark sector, chiral symmetry is
explicitly broken and therefore these interactions will not appear.

For higher momentum transfer quarks still interact through
gluon exchanges. Following de R\'{u}jula {\it et al.} \cite{ruju}
the one-gluon-exchange (OGE) interaction is taken as a 
standard color Fermi-Breit potential.
In order to obtain a unified description of light, strange and 
heavy mesons a running strong coupling constant has to be used \cite{godf}. 
The perturbative expression for $\alpha_s(Q^2)$ diverges 
when $Q\rightarrow\Lambda_{QCD}$ and therefore the coupling
constant has to be frozen at low energies. 
We parametrize this behavior by
means of an effective scale dependent strong 
coupling constant \cite{coup}
\begin{equation}
\alpha_s(\mu)={\alpha_0\over{ln\left({{\mu^2+\mu^2_0}
\over\Lambda_0^2}\right)}},
\label{asf}
\end{equation}
where $\mu$ is the reduced mass of the $q\bar q$ system and $\alpha_0$, $\mu_
0$ and $\Lambda_0$ are fitted parameters \cite{vij1}. 
This equation gives rise to $\alpha_s\sim0.54$ for the light quark sector,
a value consistent with the one used in the study of the nonstrange 
hadron phenomenology \cite{ente}, and
it also has an appropriate high $Q^2$ behavior, 
$\alpha_s\sim0.127$ at the $Z_0$ mass \cite{pre1}.
The $\delta$ function appearing in the 
OGE has to be regularized in order to avoid an unbound spectrum 
from below. Taken into account that for a coulombic system 
the typical size scales with the reduced mass, we use a 
flavor-dependent regularization $r_0(\mu)={{\hat r_0}/{\mu}}$ \cite{vij1}.
Moreover the Schr\"{o}dinger equation cannot be solved numerically 
for potentials containing $1/r^3$ terms.
This is why we regularize the noncentral terms of the
OGE with a monopole form factor with 
parameter $\Lambda _{g}(\mu )=\mu /\Lambda _{g}^{0}$ \cite{vij1}.

The other nonperturbative property of QCD is confinement.
Lattice QCD studies show that $q\overline{q}$ systems 
are well reproduced at short distances by a linear potential that
it is screened at large distances due to pair creation \cite{bali}.
One important question which has not been properly answered 
is the covariance property of confinement. While the
spin-orbit splittings in heavy quark systems 
suggest a scalar confining potential \cite{luca}, in Ref. \cite
{szcz} showed that the Dirac structure of confinement is of vector nature in
the heavy quark limit of QCD. On the other hand, a significant mixture
of vector confinement has been used to explain the decay widths 
of $P$-wave $D$ mesons \cite{suga}.
Such property being irrelevant for the central part
of the interaction, determines the sign and strength of the
spin-orbit Thomas precession term. Therefore, we write 
the spin-orbit contribution of the confining interaction 
as an arbitrary combination of scalar and vector terms
$V^{SO}_{CON}(\vec{r}_{ij})= (1-a_s) 
V^{SO}_V(\vec{r}_{ij}) + a_s V^{SO}_S(\vec{r}_{ij})$
where $V^{SO}_{V}(\vec r)[V^{SO}_{S}(\vec r)]$ is the vector (scalar)
spin-orbit contribution.

\section{Results}

With the quark-quark interaction described above we have solved
the Schr\"{o}dinger equation for the different $q\overline{q}$ systems.
Let us briefly discuss the parameters of our model. 
Most of the parameters of the Goldstone boson fields are taken 
from the $NN$ sector. The eta
and kaon cutoff masses are related with the sigma and pion one as explained
in Ref. \cite{muli}: $\Lambda [u(d)s] \simeq \Lambda(ud)+m_{s}$
where $m_{s}$ is the strange quark current mass. 
The confinement parameters $a_{c}$ and $\mu_{c}$ are fitted to
reproduce the energy difference between the $\rho$ meson and its first
radial excitation and the $J/\psi$ and the $\psi (2S)$. The
parameters involved in the OGE are obtained from
a global fit to the hyperfine splittings well established in 
the Particle Data Group (PDG) \cite{pdgb}.
Finally, one has to fix the relative strength of the scalar and
vector confinement.
Using a pure scalar confining potential one obtains 1363 MeV
and 1142 MeV for the $a_{1}(1260)$ and $a_{2}(1320)$, respectively, 
in complete disagreement with the order and magnitude 
of the experimental data. Introducing a small mixture
of vector confinement, $a_s$=0.777, the experimental
order is recovered, being now the masses 1198 MeV and 1322 MeV, respectively,
both within the experimental error bars. This value also allows to obtain
a good agreement with the experimental data in the $c\overline{c}$ and 
$b \overline{b}$ systems.

\begin{figure}[h!]
\begin{center}
\mbox{\epsfxsize=70mm\epsffile{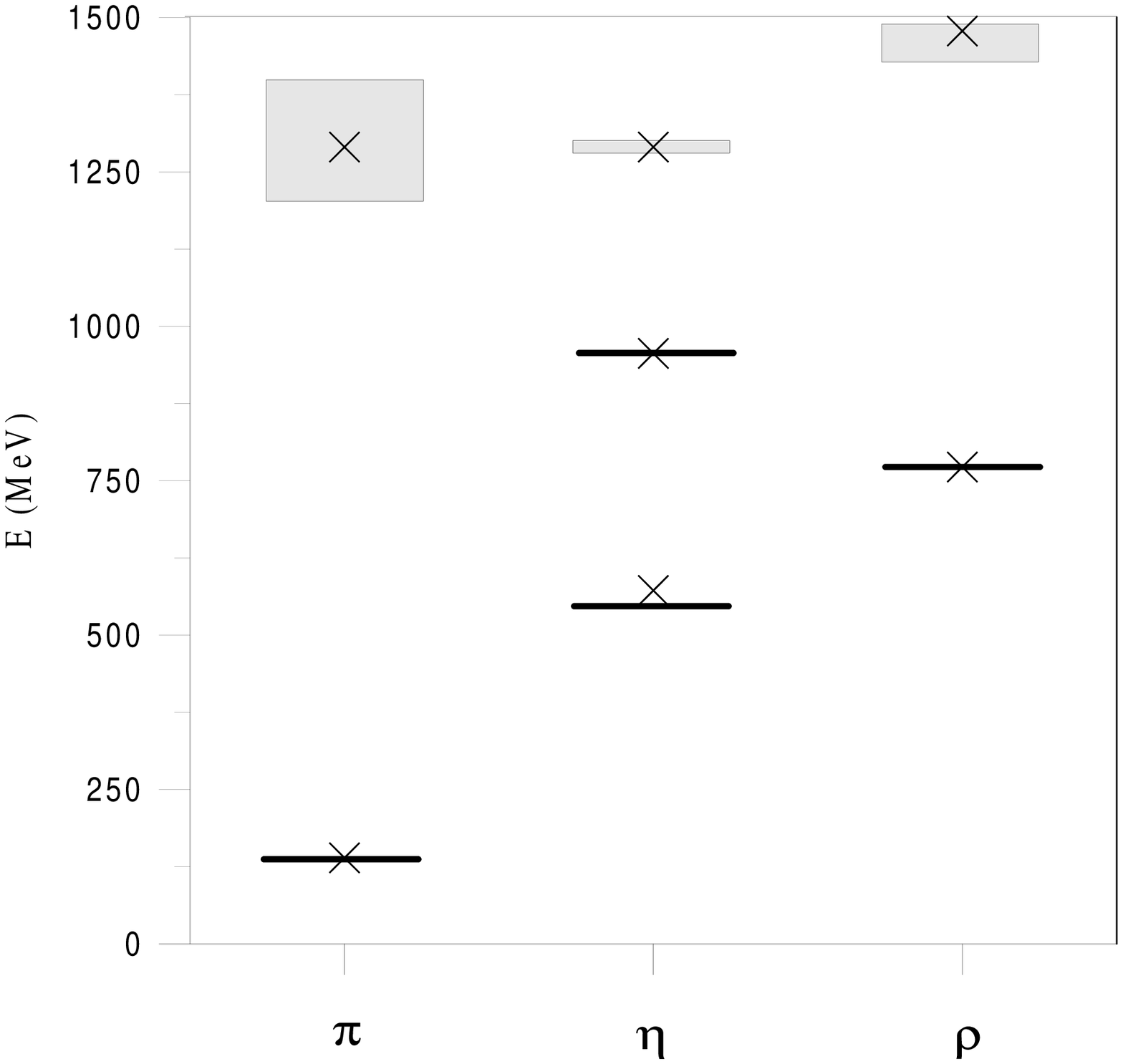}}
\mbox{\epsfxsize=70mm\epsffile{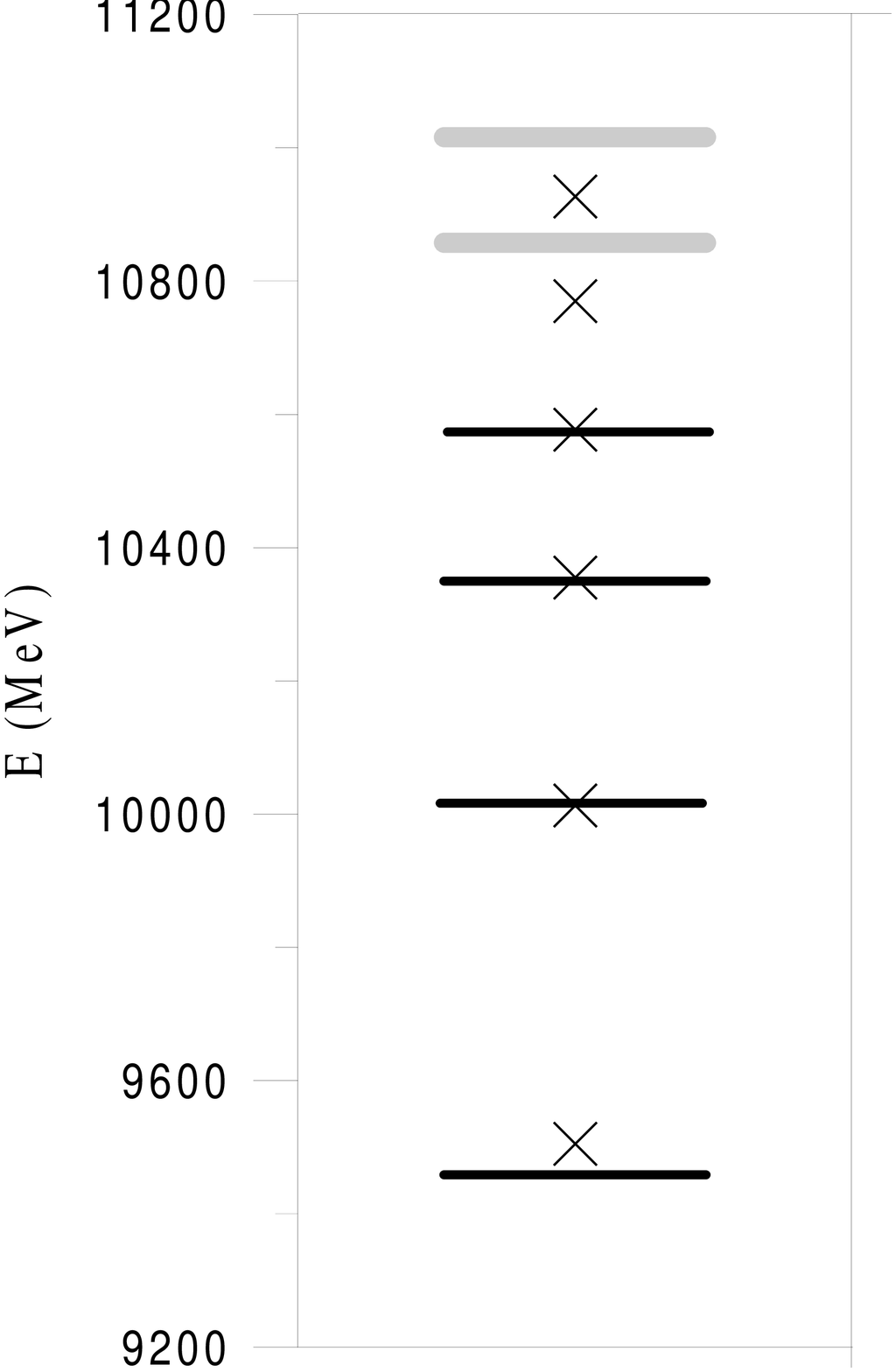}}
\end{center}
\caption{Meson spectra 
(a) light peudoscalar and vector mesons (b) $b\bar b$ mesons}
\label{fig1}
\end{figure}

In Fig. \ref{fig1} we show results for the light pseudoscalar and
vector mesons and for heavy mesons. The agreement with experimental
data is remarkable. Let us emphasize that
with only 11 parameters we are
able to describe more than 110 states \cite{vij1}. 

\subsection{The $\eta _{c}(2S)$ and $D_{s_{J}}$ states}

Recently Belle and BaBar collaborations have reported new experimental
measurements that immediately prompted different interpretations.
The Belle collaboration has reported the following
values for the mass of the $\eta _{c}(2S)$
\[
M[\eta _{c}(2S)]=\left\{ 
\begin{array}{c}
3654 \pm 6 \pm 8 \text{ MeV Ref. \cite{bel1}} \\ 
3622 \pm 6 \pm 6 \text{ MeV Ref. \cite{bel2}} \\
3630 \pm 8 \text{ MeV Ref. \cite{bel3}} 
\end{array} \right.
\]
that are significantly larger than most predictions of constituent
quark models and the previous experimental value of the PDG: 
$M[\eta_c(2S)]=3594 \pm 5$ MeV.
It has been pointed out that these values cannot be easily explained
in the framework of constituent quark models because the resulting $2S$
hyperfine splitting (HFS) would be smaller than the predicted for the $1S$
ones. In fact the predicted ratio of the $2S$ to $1S$ HFS in charmonium is
\[
R=\frac{\Delta M_{2S}^{HFS}}{\Delta M_{1S}^{HFS}}= \left\{
\begin{array}{c}
0.84 \text{ Ref. \cite{eber}} \\ 
0.67 \text{ Ref. \cite{hfs2}} \\ 
0.60 \text{ Ref. \cite{hfs3}}
\end{array} \right.
\]
whereas the experimental one is R=0.273 if $M[\eta_c(2S)]=$ 3654 MeV,
R=0.547 for 3622 MeV, and R=0.479 for 3630 MeV. 
In view of these differences some authors have claimed for
an $\alpha _{s}$ coupling constant depending on the radial excitation.

Our result is $M[\eta _c(2S)]=$ 3627 MeV, within
the error bar of the last two Belle measurements, the ones
obtained with higher statistics. Moreover the ratio 2S to 1S HFS is
found to be 0.537, in perfect agreement with the experimental data. The
reason for this agreement can be found in the shape of the 
confining potential that also influences the HFS, the
linear confinement being not enough flexible to
accommodate both excitations \cite{pedr}.

The experimental measurement reported by BaBar is a narrow
state near 2317 MeV known as $D_{S_{J}}^{\ast }(2317)$ \cite{baba}. This
state has been confirmed by CLEO \cite{cleo} together with another
possible resonance around 2460 MeV. Both
experiments interpret these resonances as $J^P=0^+$ and $1^+$
states. This discovery has triggered a series of articles \cite{arta} 
either supporting this interpretation or presenting
alternative hypothesis.

The most striking aspect of these two resonances is that their masses are
much lower than expected. 
The mass difference between the PDG mass value $D_{1}(2420)$
and the lowest value of $D_{0}^*$ reported by the Belle collaboration is
about 135 MeV. Using these data and the PDG mass value for the 
$D_{S_{1}}(2536)$ one would expect the $D_{S_{0}}^{\ast }$ mass to be 
around 2400 MeV, almost 100 MeV greater 
than the measured values. Our results are shown in Table \ref{t2}. They
agree reasonably well with the
values of the PDG for both for the $D$'s and $D_{s}$'s states,
but fail to reproduce the two $D_{s}$ states reported by Belle 
and CLEO. We also agree with the
results by FOCUS \cite{focu} for the $D_{0}$ state but we 
are far from the result of Belle for the same state. 

\section{The scalar sector: \lowercase{$q\bar q$} study}

It is still not clear which are the members of the $0^{++}$ nonet corresponding
to $L=S=1$ ${q}\overline{q}$ multiplets. 
There are too many 0$^{++}$ mesons observed in the 
region below 2 GeV to be explained as $q\overline{q}$ states. 
There have been reported in the PDG 
two isovectors $IJ^{PC}=10^{++}$:
$a_{0}(980)$ and $a_{0}(1450)$; five isoscalars $IJ^{PC}=00^{++}$: 
$f_{0}(600), f_{0}(980), f_{0}(1370), f_{0}(1500)$ and $f_{0}(1710)$;
and three $IJ^{PC}={1\over 2} 0^{++}$: 
$K_{0}^{\ast }(1430)$, $K_{0}^{\ast }(1940)$ and recently
$\kappa (900)$. The naive quark model predicts the existence
of one isovector, two isoscalars and two $I=1/2$ states. 

Our results are shown in Table \ref{t1}.
Using this table one can try to assign physical states
to $0^{++}$ nonet members. We observe that
there are no nonstrange states with $I=0$ and mass close to 1 either 1.5 GeV,
which would correspond to the $f_{0}(980)$ and the 
$f_{0}(1500)$. The same occurs in the strange sector 
around 0.8 GeV, we do not find a $q\bar{q}$ partner for the
$\kappa (900)$. 

Let us discuss each state separately. With respect to the
isovector states, there appears a candidate
for the $a_{0}(980)$, the $^3P_0$ member of the lowest $^3P_J$ 
isovector multiplet. The other candidate, the $a_{0}(1450)$, 
is predicted to be the scalar member of a $^3P_J$ excited isovector multiplet. 
This reinforces the predictions of the naive quark model, where 
the $LS$ force makes lighter the $J=0$ states with respect to the $J=2$.
The assignment of the $a_0(1450)$ 
as the scalar member of the lowest $^3P_J$ multiplet
would contradict this idea, because the $a_2(1312)$ is well established
as a $q \bar q$ pair. The same behavior is evident in the
$c\overline{c}$ and the $b\overline{b}$ spectra, making impossible to
describe the $a_{0}(1450)$ as a member of the lowest $^3P_J$
isovector multiplet without spoiling the description 
of heavy-quark multiplets.
However, in spite of the correct description of the mass 
of the $a_{0}(980)$, the model predicts a pure light-quark content,
what seems to contradict some experimental evidences. 
The $a_{0}(1450)$ is predicted to be also a pure light quark structure
obtaining a mass somewhat higher than the experiment.

In the case of the isoscalar states, one finds a candidate for the $f_{0}(600)$
with a mass of 402 MeV, in the lower limit of the
experimental error bar and with a strangeness content around 
$8\%$. The $f_{0}(980)$ and $f_{0}(1500)$ cannot be found for any
combination of the parameters of the model. 
It seems that a different structure rather
than a naive $q\overline{q}$ pair is needed to describe
these states. The $f_0(1500)$ is a clear
candidate for the lightest glueball \cite{amsl} and our results support this
assumption. Concerning the $f_{0}(1370)$ (which may actually correspond to
two different states \cite{blac}) we obtain two almost degenerate states
around this energy, the lower one with a predominantly nonstrange content,
and the other with a high $s\overline{s}$ content. 
Finally a state corresponding to the $f_{0}(1710)$ is obtained.

Concerning the $I=1/2$ sector, as a consequence of the larger mass of the
strange quark as compared to the light ones, our model always predicts a mass
for the lowest $0^{++}$ state 200 MeV greater than the 
$a_{0}(980)$ mass. Therefore, being the $a_0(980)$ the member of the
lowest isovector scalar multiplet, the $\kappa (900)$
cannot be explained  as a $q\bar q$ pair.
We find a candidate for the $K_0^*(1430)$ although with a smaller mass.

In conclusion our results indicate that the light scalar sector cannot be
described in a pure $q\overline{q}$ scheme and more complicated
structures or mixing with multiquark states seems to be needed. 
Our conclusion concerning the 
$f_{0}(980)$ and the $\kappa (900)$ are very similar to the ones obtained
by Umekawa \cite{umek} using the extended Nambu-Jona-Lasinio model in an
improved ladder approximation of the Bethe-Salpeter equation. This fact
seems to indicate that relativistic 
corrections would not improve the situation and
the conclusions remain model independent.

\section{The scalar sector: tetraquark study}
 
In the naive quark model, to construct
a positive parity state requires a unit of angular momentum
in a $q \bar q$ pair. Apparently, this takes an 
energy around 1 GeV since similar
meson states ($1^{++}$ and $2^{++}$) lie above 1.2 GeV.  
However a more complicated structure, like $q^2 \bar q^2$, suggested
twenty years ago by Jaffe \cite{jaff} 
can couple to $0^{++}$ without orbital excitation
and therefore could be a serious 
candidate to explain the structure of some light scalar mesons.

In this section we study tetraquark bound states, focusing
our attention in those states with the quantum 
numbers of the scalar mesons. 
We solve the Schr\"odinger equation 
using a variational method where the spatial trial
wave function is a linear combination of gaussians 
The technical details are given
in Ref. \cite{vij2}. Let us only mention that neglecting the
exchange terms in the variational wave function
is fully justified for the heavy-light tetraquarks
but it may induce some corrections in the present study.
Due to the presence of the kaon-exchange
there is a mixture among different configurations
with the same isospin. In particular, in the isoscalar sector
the configurations: $[(qq)(\bar q \bar q)]$,
$[(qs)(\bar q \bar s)]$, and
$[(ss)(\bar s \bar s)]$ are mixed. The same happens in the isovector
case for the configurations:
$[(qq)(\bar q \bar q)]$, and
$[(qs)(\bar q \bar s)]$, and in the $I=1/2$ case for the configurations:
$[(qq)(\bar q \bar s)]$, and
$[(qs)(\bar s \bar s)]$. In all cases $q$ stands for a $u$ or $d$
quark.

\begin{figure}[h!]
\begin{center}
\mbox{\epsfxsize=70mm\epsffile{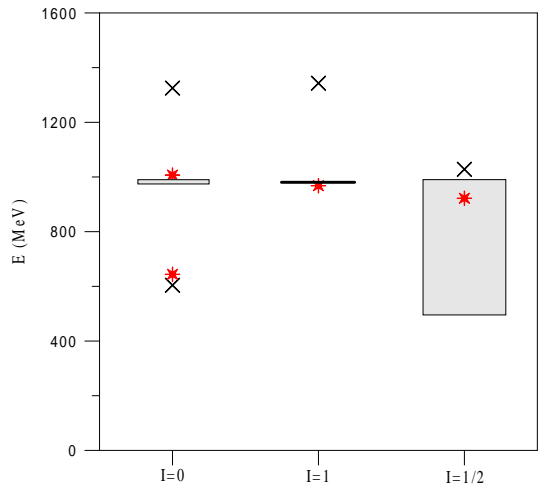}}
\end{center}
\caption{Tetraquark masses for the scalar sector}
\label{fig2}
\end{figure}

The results are shown in Fig. \ref{fig2}. We present the lowest
states for the three isospin sectors. The $\times$'s show 
the results obtained within the model described in 
Sect. II. As one can see, there appear two states, in the
isoscalar and isovector sectors, with almost the same mass, 
although too high to be identified with the $f_0(980)$
and $a_0(980)$. In the $I=1/2$ sector, there appears a
candidate to be identified with the $\kappa(900)$. 
It has been recently argued the possible importance of 
three-body forces arising from the confining interaction 
for those systems containing at least three quarks \cite{dimi}.
We have performed a calculation including a three-body confining
term as the one reported in Ref. \cite{dimi}. The strength of 
this interaction has been chosen to reproduce the mass of the 
$f_0(980)$. The results are plotted
in Fig. \ref{fig2} by $\star$'s. As one can see the
degeneracy between the isoscalar and isovector states remains, while
the lowest state of the isoscalar and $I=1/2$ sectors are almost 
not affected. From these results one could inferred that the
scalar sector needs the presence of tetraquark structures
to be understood. 

As a summary, we have obtained candidates for all the light scalar mesons 
reported in the literature. Our results suggest
that there would be some states, as it is the case of the
$a_0(980)$ and $f_0(600)$ that would be a mixture of
a $q\bar q$ and tetraquark structure, but it definitively
assigns a tetraquark structure to the 
$f_0(980)$ and the $\kappa(900)$.

\section{ acknowledgments}

This work has been partially funded by Ministerio de 
Ciencia y Tecnolog{\'{\i}}a under Contract No. BFM2001-3563
and by Junta de Castilla y Le\'{o}n under
Contract No. SA-109/01.

\begin{table}[tbp]
\caption{$D$'s and $D_s$'s masses in MeV compared to all known experimental
data \protect\cite{bian}}
\label{t2}
\begin{center}
\begin{tabular}{ccccccc}
\hline
& $M^*(0^+)$ & $M(1^+)$ & $M(1^+)$ & $M^*(2^+)$ & $%
M(0^-)(2S)$ & $M^*(1^-)(2S)$ \\ \hline
\multicolumn{7}{c}{$D$'s} \\ \hline
PDG (0) &  &  & 2422$\pm$2 & 2459$\pm$2 &  &  \\ 
PDG ($\pm$) &  &  & 2427$\pm$5 & 2459$\pm$4 & 2637$\pm$7 & 2637$\pm$7 \\ 
FOCUS (0) & $\sim 2420$ &  &  & 2463$\pm$2 &  &  \\ 
FOCUS ($\pm$) & $\sim 2420$ &  &  & 2468$\pm$2 &  &  \\ 
Belle (0) & 2290$\pm$30 & 2400$\pm$36 & 2424$\pm$2 & 2461$\pm$4 &  &  \\ 
CLEO (0) &  & 2461$\pm$51 &  &  &  &  \\ 
&  &  &  &  &  &  \\ 
Our result & 2437.9 & 2496.0 & 2495.7 & 2500.5 & 2641.7 & 2700.1 \\ 
&  &  &  &  &  &  \\ \hline
\multicolumn{7}{c}{$D_s$'s} \\ \hline
PDG ($\pm$) &  &  & 2535.3$\pm$0.6 & 2572.4$\pm$1.5 &  &  \\ 
FOCUS ($\pm$) &  &  & 2535.1$\pm$0.3 & 2567.3$\pm$1.4 &  &  \\ 
BaBar & 2317 &  &  &  &  &  \\ 
CLEO & 2317 & 2460 &  &  &  &  \\ 
&  &  &  &  &  &  \\ 
Our result & 2470.5 & 2565.5 & 2549.7 & 2584.4 & 2698.7 & 2764.3 \\ 
&  &  &  &  &  &  \\ \hline
\end{tabular}
\end{center}
\end{table}

\begin{table}[h!]
\caption{Light scalar meson masses in MeV}
\label{t1}
\begin{center}
\begin{tabular}{cccc}
\hline
State $(n^{2I+1,2S+1}L_J)$ & Meson & Our result & Experiment \\ 
\hline
$1^{3,3}P_0$ & $a_0(980)$ & 983.5 & 984.7$\pm$1.2 \\ 
$2^{3,3}P_0$ & $a_0(1450)$ & 1586.3 & 1474$\pm$19 \\ 
$1^{1,3}P_0$ & $f_0(600)$ & 402.7 & 400$-$1200 \\ 
$1^{1,3}P_0$ & $f_0(1370)$ & 1341.7 & 1200-1500 \\ 
$2^{1,3}P_0$ & $f_0(1370)$ & 1391.2 & 1200-1500 \\ 
$2^{1,3}P_0$ & $f_0(1710)$ & 1751.8 & 1713$\pm$6 \\ 
$3^{1,3}P_0$ & $f_0(2020)$ & 1893.8 & 1992$\pm$16 \\ 
$3^{1,3}P_0$ & $f_0(2200)$ & 2212.2 & 2197$\pm$11 \\ 
$1^{2,3}P_0$ & $K^*_0(1430)$ & 1213.5 & 1412$\pm$6 \\ 
$2^{2,3}P_0$ & $K^*_0(1950)$ & 1768.5 & 1945$\pm$30 \\ \hline
\end{tabular}
\end{center}
\end{table}

\end{document}